\documentclass{article}
\usepackage{graphicx}
\pdfoutput=1
\usepackage{amsmath, amssymb}
\usepackage{amsthm}
\usepackage{amsfonts}
\usepackage{subfigure}

\newcommand\noi{\noindent}

\newcommand\xc{X_C}
\newcommand\xe{X_E}

\newcommand\cs{\mathcal{S}}
\numberwithin{equation}{section}
\theoremstyle{plain}
\newtheorem{thm}{Theorem}
\newtheorem{lem}[thm]{Lemma}
\newtheorem{prop}[thm]{Proposition}

\newtheorem{assertion}[equation]{}
\theoremstyle{definition}
\newtheorem{notation}[equation]{}
\newtheorem*{defn}{Definition}

\begin{document}

\title{Configuration spaces of convex and embedded polygons in the plane}
\author{Don Shimamoto\footnote{Department of Mathematics and Statistics, Swarthmore College, 500 College Ave., Swarthmore, PA 19081. \texttt{dshimam1@swarthmore.edu}.} \ and Mary Wootters\footnote{Department of Mathematics and Statistics, Swarthmore College, 500 College Ave., Swarthmore, PA 19081. \texttt{mary.wootters@gmail.com}.  Supported by a Swarthmore College Summer Research Fellowship.}}
\date{}

\maketitle

A celebrated result of Connelly, Demaine, and Rote [\textbf{6}] states that any polygon in the plane can be ``convexified."  That is, the polygon can be deformed in a continuous manner until it becomes convex, all the while preserving the lengths of the sides and without allowing the sides to intersect one another.  In the language of topology, their argument shows that the configuration space of \textsl{embedded} polygons with prescribed side lengths deformation retracts onto the configuration space of \textsl{convex} polygons having those side lengths.  In particular, both configuration spaces have the same homotopy type.  Connelly, Demaine, and Rote observe (without proof) that the space of convex configurations is contractible.  Separately, work of Lenhart and Whitesides [\textbf{10}] and of Aichholzer, Demaine, Erickson, Hurtado, Overmars, Soss, and Toussaint [\textbf{1}] had shown that the space of convex configurations is connected.  These results are part of  the literature on linkages.  The polygons here are mechanical linkages in which the sides can be viewed as rigid bars of fixed length arranged in a cycle and the vertices as joints  about which the bars can rotate.

In this note we determine the topology of  the space of convex configurations and the space of embedded configurations up to homeomorphism.   We regard two polygons as equivalent if one can be translated and rotated onto the other. After a translation, we may assume that one of the vertices is at the origin and then, after a rotation, that one of the adjacent sides lies along the positive $x$-axis.  To fix some notation, let $\vec{\ell} = (\ell_1, \dots , \ell_n)$ be a given sequence of side lengths, where $\ell_i >0$ for all $i$ and $n \geq 3$.  The configuration space of planar polygons with these side lengths is defined by:
\begin{equation*}
\begin{split}
X(\vec{\ell}) = \{ (p_1, \dots , p_n) & \in({\bf R}^2)^n \; :  \;  p_1 = (\ell_1, 0), p_n = (0,0), \hbox{ and }\\
&  |p_i - p_{i-1}| = \ell_i  \hbox{ for } i= 1, \dots, n \}.
\end{split}
\end{equation*}
 Here, and in what follows, subscripts should be taken modulo $n$ where appropriate.  Also, to simplify the notation, we suppress the dependence on $\vec{\ell}$, writing $X$ rather than $X(\vec{\ell})$. Note that $X$ inherits a natural topology as a subspace of $({\bf R}^2)^n$. To any element $p = (p_1, \dots , p_n)$ in $X$ we associate its sequence of {\it turn angles} $\theta = (\theta_1, \dots , \theta_n)$, where $\theta_i$ is the signed angle at $p_i$ from the $i$th edge $e_i = p_i - p_{i-1}$ to the $(i + 1)$th edge $e_{i+1} = p_{i+1} - p_i$.  We choose $\theta_i$ to lie in the interval $(- \pi , \pi]$.  A positive value of $\theta_i$ corresponds to a left turn, and a negative value to a right turn.
 
\begin{figure}[htb] 
\vskip -1.5 in
\hskip -.25 in \includegraphics[width=7.5in]{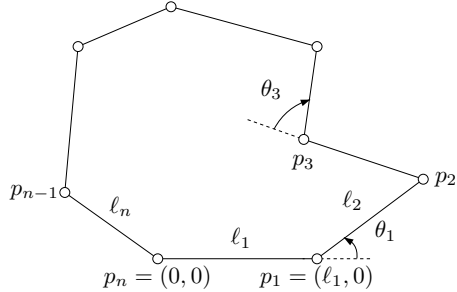}
\vskip -6.5 in
\caption{A polygon in the plane with vertices, side lengths, and turn angles labeled.}

\label{fig:polygon}
\end{figure}

The definition of $X$ permits polygons that intersect themselves.  We exclude such configurations by focusing  on embedded polygons, that is, those in which the edges do not intersect except at common endpoints.  These polygons fall into two components depending on whether they are traversed counterclockwise ($\sum_i \theta_i = 2 \pi$) or clockwise ($\sum\theta_i = -2\pi$).  The two components are homeomorphic to one another by the transformation that sends each polygon to its reflection in the $x$-axis.  Here we concentrate on the counterclockwise component: 

\begin{notation}
\textsl{Let $\xe$ denote the set of polygons $p = (p_1, \dots , p_n)$ in $X$ that are embedded  in ${\bf R}^2$ and satisfy $\sum_i \theta_i = 2 \pi.$}
\end{notation} 

Within $\xe$ is the subset of convex configurations: 

\begin{notation}
\textsl{Let $\xc$ denote the subset of $\xe$ consisting of convex polygons. }
\end{notation}

\medskip

\noi These are the elements of $\xe$ whose turn angles satisfy $\theta_i \geq 0$ for all $i$.

The topology of the configuration space $X$ has been studied using a variety of techniques by several authors (for instance, Hausmann [\textbf{7}],  Kamiyama [\textbf{8}], and Kapovich and Millson [\textbf{9}]).  The analysis can get complicated, but certain basic properties are easy to establish.  For instance, suppose that $X$ contains no straight line configurations, that is, no configurations in which all the edges  lie along a single line.  This is the generic situation, since $X$ contains a straight line configuration if and only if it is possible to choose values $\epsilon_i = \pm 1$ such that $\sum_i \epsilon_i \ell_i = 0$.   Then it is well known that $X$ is  a $C^\infty$ compact orientable manifold of dimension $n - 3$ (see Shimamoto and Vanderwaart [\textbf{12}] for a recent exposition).  The space of embedded polygons $\xe$ is an open subset of $X$, hence also an $(n-3)$-dimensional manifold.  The space of convex configurations $\xc$ is a closed subset of $X$ (though not a smooth one, as we shall see).  Here it is still important to exclude straight line configurations, since otherwise a sequence of convex configurations could converge to a straight line configuration, which would be a limit point not in $\xc$.

For the remainder of the paper, assume that the side lengths are such that $X$ contains no straight line configurations.  Our main results describe the topological type of $\xc$ and $\xe$.  Specifically, we prove that:

\begin{assertion}
 $\xc$ is homeomorphic to a closed $(n-3)$-dimensional ball $B$, e.g., to $B = \{ x \in {\bf R}^{n-3} : | x | \leq 1 \}$,
 \end{assertion}
 
and

\begin{assertion}
$\xe$ is homeomorphic to ${\bf R}^{n-3}$.
\end{assertion}

\medskip

Statement (0.3) is proved in section 1 (where it is called Theorem 5), and (0.4) is proved in section 2 (Theorem 7).  We close in section 3 with a counterexample to a conjecture of Connelly, Demaine, and Rote regarding the closure of $\xe$ in $X$.

\section{Convex configurations}

A polygon $p = (p_1, \dots, p_n)$ in $X$ has a sequence of turn angles $\theta = (\theta_1, \dots, \theta_n)$ in ${\bf R}^n$.  But conversely the turn angles determine the polygon as well, since $p_1 = (\ell_1, 0)$ by our convention  and, for $j> 1$, 
\[
p_j = p_{j-1} + \ell_j \bigl(\, \cos(\sum_{k=1}^{j-1} \theta_k) \, ,\,  \sin(\sum_{k=1}^{j-1} \theta_k) \, \bigr).  
\]
 In this section we will be concerned only with convex polygons, in which case $\theta_i \in [0 , \pi)$ for all $i$.  This eliminates any worries about continuity problems modulo $2 \pi$ as the angles vary.  Thus the function $t (p_1, \dots, p_n) = (\theta_1, \dots, \theta_n)$ that sends a polygon to its turn angles maps the subspace $\xc$ of convex configurations homeomorphically onto its image $\cs = t(\xc)$ in ${\bf R}^n$.  The set $\cs$ consists of all sequences of turn angles that are realized by convex polygons with side lengths $\vec{\ell}$.  We will determine the topological type of $\xc$ by studying how it is parametrized by $\cs$, essentially executing a search of all possible turn angles.  Actually, a convex polygon is determined by its first $n-3$ turn angles alone, but for our arguments it is convenient to keep track of all $n$ of them.

 The general strategy here is to examine inductively how much freedom there is to rotate each of the edges.  For instance, the first edge is fixed from $p_0 = p_n = (0,0)$ to $p_1 = (\ell_1, 0)$ by our convention.  We then see how much the second edge can be ``wiggled" under the restriction that the whole polygon remains convex.  For each position reached by this motion, we see how much the next edge can be wiggled, and so forth.  By doing this for all the edges (or at least for the first $n-2$ of them), all configurations will have been visited.  We show that each of the wiggle ranges is an interval, which is nontrivial except possibly when one of the previous positions was set to an interval endpoint.  Moreover, the endpoints of these intervals vary continuously with the angle choices in the positions that precede it.    Thus the space of convex configurations can be built up iteratively as a union of segments of continuously varying length.

\subsection{The maximum and minimum turn angles}  

For $k \leq n$, let $\pi_k \colon {\bf R}^n \to {\bf R}^k$ be the projection onto the first $k$ coordinates, $\pi_k(x_1, \dots, x_n) = (x_1, \dots , x_k)$, and let $\cs_k = \pi_k(\cs)$.    In words,
\begin{equation*}
\begin{split}
\cs_k = &\{ (\theta_1, \dots , \theta_k) \> : \> \hbox{there exist } \theta_{k+1} , \dots , \theta_n \hbox{ such that } (\theta_1, \dots , \theta_n) \hbox{ is the}\\ 
 &\hbox{sequence of turn angles of a convex polygon } p \}. \\
\end{split}
\end{equation*}
We determine the topology of $\cs$ by analyzing the relation between $\cs_k$ and $\cs_{k-1}$ inductively.

 For example, $\cs \approx \xc$ is connected [\textbf{1}], [\textbf{10}] and compact, so, since $\pi_1$ is continuous,  $\cs_1$ is a finite closed interval, say 
\begin{equation}
 \cs_1 = [\nu_1, \mu_1]. 
 \end{equation}
 Here $\mu_1$ represents the maximum possible turn angle at $p_1$, and $\nu_1$ represents the minimum.

In general, if $k>1$, let $\alpha = (\alpha_1, \dots , \alpha_{k-1})$ in $\cs_{k-1}$ be given.  These turn angles determine a fixed chain of $k$ edges from $p_0$ to $p_k$.  Define
\[
R_k(\alpha) = \{ \theta_k \in {\bf R}  \> : \> (\alpha, \theta_k) \in \cs_k \}.
\]
This represents the wiggle room mentioned earlier for the $(k+1)$th edge, given fixed positions for the first $k$ edges.  Next let
\[
\nu_k(\alpha) = \inf R_k(\alpha) \qquad \hbox{and} \qquad \mu_k(\alpha) = \sup R_k(\alpha).
\]
These are the smallest and largest possible turn angles at $p_k$, again assuming an initial fixed chain up to that point is given.

We will describe various properties of $\nu_k$ and $\mu_k$ including the fact that they are actually attained, that is, $\nu_k(\alpha), \mu_k(\alpha) \in R_k(\alpha)$, by deforming our polygons into certain standard configurations.   In some cases, we sketch details of proofs when the geometry is clear, making particular use of the observations in Aichholzer et al.~[\textbf{1}], for instance:

\begin{lem}  Given a convex quadrilateral with vertices $v_1, v_2, v_3, v_4$, there is a motion that increases the turn angles at $v_1$ and $v_3$ and decreases the turn angles at $v_2$ and $v_4$ while preserving the lengths of the sides.  The motion can continue until one of the  turn angles reaches $0$ or $\pi$.
\end{lem}

\begin{figure}[htb]
\vskip -1.75 in
\hskip -.75 in \includegraphics[width=7.5in]{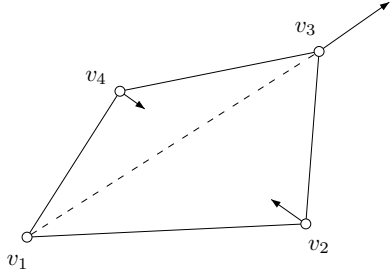}
\vskip -6.5 in
\caption{Increasing the turn angles at $v_1$ and $v_3$ and decreasing the turn angles at $v_2$ and $v_4$.}
\end{figure}

The idea is to move $v_3$ directly away from $v_1$ along the ray from $v_1$ to $v_3$.  (See Figure 2.) The positions of $v_2$ and $v_4$ are then determined by the side lengths.    The authors in [\textbf{1}] use this result to show that, given any two convex polygons with the same sequence of edge lengths, there is a continuous motion from one to the other in which all angles change monotonically.  The monotonicity ensures that the intermediate polygons are also convex, whence the space $\xc$ is connected.

For convenience, we introduce the following terminology.

\begin{defn} 
In a polygon $ (p_1, \dots , p_n)$ with turn angles $(\theta_1, \dots, \theta_n)$, a vertex $p_i$ is called {\it flat} if $\theta_i = 0$.
\end{defn}

Now consider the minimum turn angle $\nu_k(\alpha)$.  If there exists a polygon having turn angles $(\alpha, \theta_k, \dots , \theta_n)$ with $\theta_k=0$, then $\nu_k(\alpha) = 0$, since that's always the smallest possible turn angle for a convex polygon.   Let us call this case (a).  Otherwise, suppose that $\theta_k > 0$.  To decrease this turn angle, we try to rotate the $(k + 1)$th edge clockwise.  If the subchain from $p_{k+1}$ to $p_n$ is not straight, choose a vertex $p_j$ along the way such that the turn angle $\theta_j$ is nonzero,   Now consider the quadrilateral $p_np_kp_{k+1}p_j$.  By keeping the subchains of the original polygon between these vertices rigid and moving $p_{k+1}$ directly away from $p_n$ as described above, we reduce the turn angle $\theta_k$.  (See Figure 3(i).  In the figure, as the inscribed quadrilateral flattens, the subchains between its vertices will  rotate.  Strictly speaking, this violates our convention that edge $p_np_1$ lie along the positive $x$-axis, so one should imagine a simultaneous compensating global rotation that keeps $p_np_1$ horizontal.)  Continue until either $\theta_k = 0$ or one of the vertices along the subchain becomes flat.  In the case of the latter, if the subchain from $p_{k+1}$ to $p_n$ is still not straight, repeat.  In this way, we eventually reach a configuration in which either $\theta_k = 0$ (case (a) again, as in Figure 3(ii)) or the subchain from $p_{k+1}$ to $p_n$ is straight (call this case (b), shown in Figure 3(iii)).  We refer to a configuration of either of these types as \textit{minimally stretched}.  (For case (a), there is not a unique such configuration, but this does not matter for our arguments.)  The procedure just given shows that, in a minimally stretched configuration, $\theta_k = \nu_k(\alpha)$, since the turn angle $\theta_k$ of any other polygon can be reduced until it reaches such a configuration.  This also implies that $R_k(\alpha)$ is connected.  

\begin{figure}[htb]
\vskip -1.5 in
\hskip - 1.25 in \includegraphics[width=7.5in]{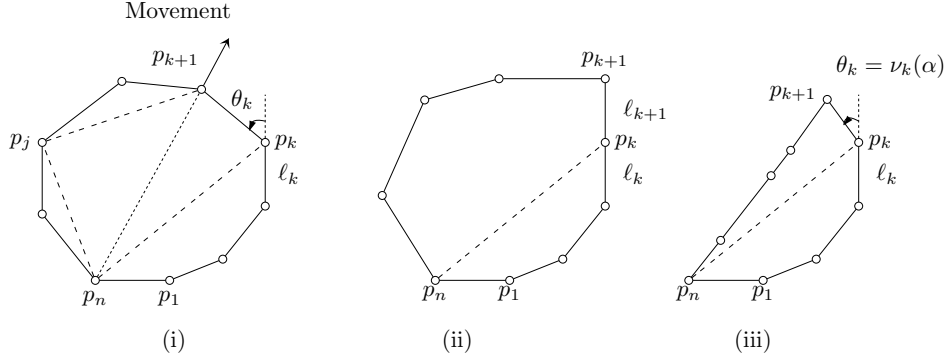}
\vskip -6.25 in
\caption{(i) Reducing $\theta_k$; (ii) case (a): $\nu_k(\alpha) = 0$; (iii) case (b): $\nu_k(\alpha) > 0$}
\end{figure}

Finally we claim that $\nu_k(\alpha)$ is a continuous function of $\alpha$, that is, a small change in the initial subchain produces a small change in the minimum turn angle.  This is clear except possibly at the overlap of cases (a) and (b), that, is, when $\nu_k(\alpha) = 0$ and the subchain from $p_{k+1}$ to $p_n$ is straight (Figure 4).  Here a small change in $\alpha$ can lead either to case (a) or (b), but either way the minimally stretched configuration changes only slightly, and hence so does the minimum turn angle.

\begin{figure}[htb] 
\vskip -1.5 in
\hskip -.5 in \includegraphics[width=7.5in]{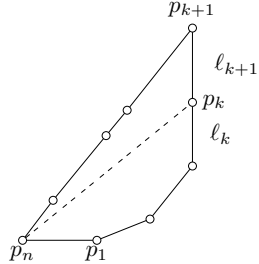}
\vskip -6.75 in
\caption{Borderline case.}
\end{figure}

The case of the maximum turn angle $\mu(\alpha)$ is similar, only now we want to rotate the $(k+1)$th edge counterclockwise as far as possible.  We do this by trying to straighten the back end of the polygon as much as we can, pulling the edge towards it in the process. To make this more precise, let a configuration with turn angles $(\alpha, \theta_k, \dots , \theta_n)$ be given.  If the subchain from $p_k$ to $p_{n-1}$ is not straight, choose a vertex $p_j$ along the way for which $\theta_j \neq 0$.   Now consider the quadrilateral $p_np_kp_jp_{n-1}$.  By moving $p_{n-1}$ directly away from $p_k$, we increase the turn angle $\theta_k$ (Figure 5(i)).  Continue until, in the original polygon, either one of the vertices along the subchain becomes flat or $p_n$ becomes flat.  In the first case, if the subchain from $p_k$ to $p_{n-1}$ is still not straight, repeat the procedure.  On the other hand, if $p_n$ is flat, repeat with $p_{n-1}$ in place of $p_n$ and look at the subchain from $p_k$ to $p_{n-2}$.   Eventually, one reaches a configuration in which the subchain from $p_k$ to $p_j$ is straight for some $j > k$ and the vertices from $p_{j+2}$ to $p_n$, if any, are flat.  Let us call such a configuration {\it maximally stretched} (Figure 5(ii)).  In particular, at most two vertices from $p_{k+ 1}$ to $p_n$ are not flat, and, if there are two, they are adjacent to one another.  

\begin{figure}[htb] 
\vskip -1.6 in
\hskip -1.5 in
\includegraphics[width=7.5in]{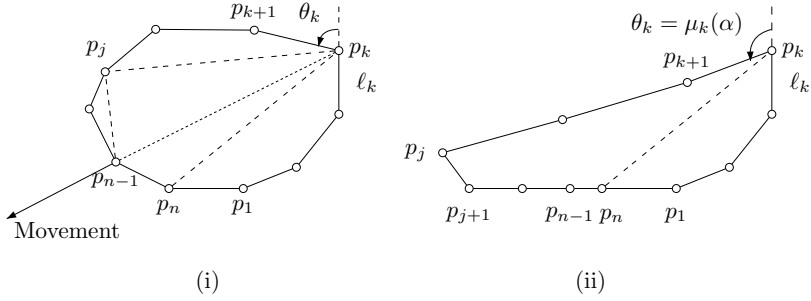}
\vskip -6.5 in
\caption{(i) Increasing $\theta_k$; (ii) maximally stretched configuration.}
\end{figure}

As with $\nu_k$, in a maximally stretched configuration $\theta_k = \mu_k(\alpha)$ because any polygon can be deformed to such a configuration with $\theta_k$ increasing along the way.  In addition, $\mu_k(\alpha)$ varies continuously with $\alpha$.  This time, the borderline case is when the subchains from $p_k$ to $p_j$ and from $p_j$ to $p_1$ are both straight.  When this happens, a small change in $\alpha$ can lead to maximally stretched configurations with straight subchains either from $p_k$ to $p_j$ or from $p_k$ to $p_{j-1}$.  While this description may sound discontinuous, the maximally stretched configurations themselves change only slightly, and hence so does $\mu_k(\alpha)$.

To recap, we have shown that $R_k(\alpha) = [\nu_k(\alpha), \mu_k(\alpha)]$.  Since by definition, $\cs_k = \{(\alpha, \theta_k) \in {\bf R}^k \> | \> \alpha \in \cs_{k-1}, \theta_k \in R_k(\alpha) \}$, the preceding discussion can be summarized as follows.
\
\begin{lem}
There exist real numbers $\nu_1$ and $\mu_1$ such that $\cs_1 = [\nu_1, \mu_1]$, and, for $k > 1$, there exist continuous  functions $\nu_k, \mu_k \colon \cs_{k-1} \to {\bf R}$ such that

\begin{equation}
\cs_k = \{ (\alpha , \theta_k) \in {\bf R}^k \> : \>  \alpha \in \cs_{k-1} , \nu_k(\alpha) \leq \theta_k \leq \mu_k(\alpha) \}. 
\end{equation}
\end{lem}

In other words, $\cs_k$ is the region over $\cs_{k-1}$ lying between two continuous graphs.

\subsection{The topology of $\cs_k$.}  
To complete the description of the topology of $\cs_k$, we use one final technical lemma.

\begin{lem}  
Suppose that $2 \leq k \leq n - 3$.  If $\alpha \in \hbox{\rm Int}\, \cs_{k-1}$, then $\nu_k(\alpha) < \mu_k(\alpha)$.
\end{lem}
 
Here the interior of $\cs_{k-1}$ is as a subset of ${\bf R}^{k-1}$.

\begin{proof}
Suppose to the contrary that $\nu_k(\alpha) = \mu_k(\alpha)$.  In other words, given the initial subchain from $p_0$ to $p_k$, the polygon is completely rigid.  Thus there exists a polygon with turn angles $(\alpha, \theta_k , \dots, \theta_n)$ that is simultaneously minimally and maximally stretched.  We consider two cases.  

First, suppose that $\theta_k = 0$ (Figure 6(i)).  Then the straight subchain from $p_k$ to $p_j$ that is part of a maximally stretched configuration is contained in a straight subchain from $p_{k-1}$ to $p_j$.  Hence, this configuration is maximally stretched at $p_{k-1}$ as well, that is, $\alpha_{k-1} = \mu_{k-1}(\alpha_1, \dots , \alpha_{k-2})$.  By $(1.2)$, this contradicts the assumption that $\alpha \in \hbox{Int}\, \cs_{k-1}$.

\begin{figure}[htb] 
\vskip -1.4 in
\hskip -1.5 in
\includegraphics[width=7.5in]{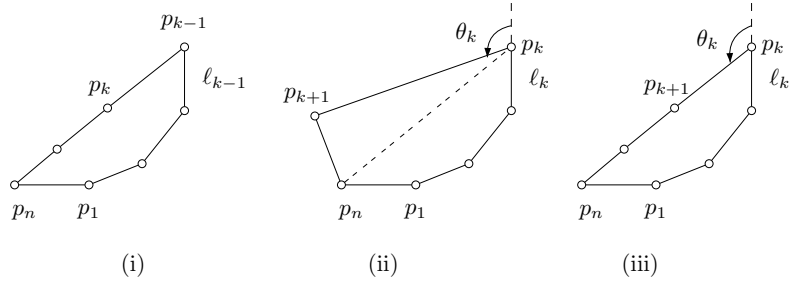}
\vskip -6.7 in
\caption{$\theta_k= \nu_k(\alpha) = \mu_k(\alpha)$: (i) $\theta_k = 0$; (ii) $\theta_k > 0$, $p_{k+1}$ not flat; (iii) $\theta_k > 0$, $p_{k+1}$ flat.}
\end{figure}

On the other hand, suppose that $\theta_k >0$.  Then, as a minimally streched configuration, the subchain from $p_{k+1}$ to $p_n$ is straight.  If $p_{k+1}$ is not flat  (Figure 6(ii)), then, as part of a maximally stretched configuration, this subchain can consist of at most one edge.  Thus $k \geq n-2$, contrary to assumption.  But if $p_{k+1}$ is flat, then the entire subchain from $p_k$ to $p_n$ is straight, meaning that the configuration is minimally stretched at $p_{k-1}$,   i.e. $\alpha_{k-1} = \nu (\alpha_1, \dots , \alpha_{k-2})$, again contradicting $\alpha \in \hbox{Int}\, \cs_{k-1}$ (Figure 6(iii)). 

In all cases we reach a contradiction.
\end{proof}

From this we obtain the topological type of $\cs_k$.

\begin{prop} 
If $1 \leq k \leq n-3$, then $\cs_k$ is homeomorphic to a closed $k$-dimensional Euclidean ball.
\end{prop}

\begin{proof}
We use induction on $k$.  For $k = 1$, the result is given by $(1.1)$.  If $k > 1$, we use $(1.2)$.  By induction, we may replace $\cs_{k-1}$ by a closed $(k-1)$-dimensional ball, up to homeomorphism  (Figure 7(i)).  Let $\partial \cs_{k-1}$ denote the boundary of $\cs_{k-1}$, a topological $(k-2)$-dimensional sphere.   Let $A$ denote the graph of $\nu_k$ restricted to $\partial \cs_{k-1}$, i.e., $A = \{ (\alpha, \nu_k(\alpha)) \> : \> \alpha \in \partial \cs_{k-1} \}$.  Similarly, let $B$ denote the graph of $\mu_k$ restricted to $\partial \cs_{k-1}$.  Choose some large number $M > 0$, and construct two cones,  $C_1$ from $A$ to the point $(0, \dots , 0 , -M) \in {\bf R}^k$ and $C_2$ from $B$ to the point $(0, \dots, 0, M)$ (Figure 7(ii)). 

\begin{figure}[htb] \centering
\includegraphics[width=4.75in]{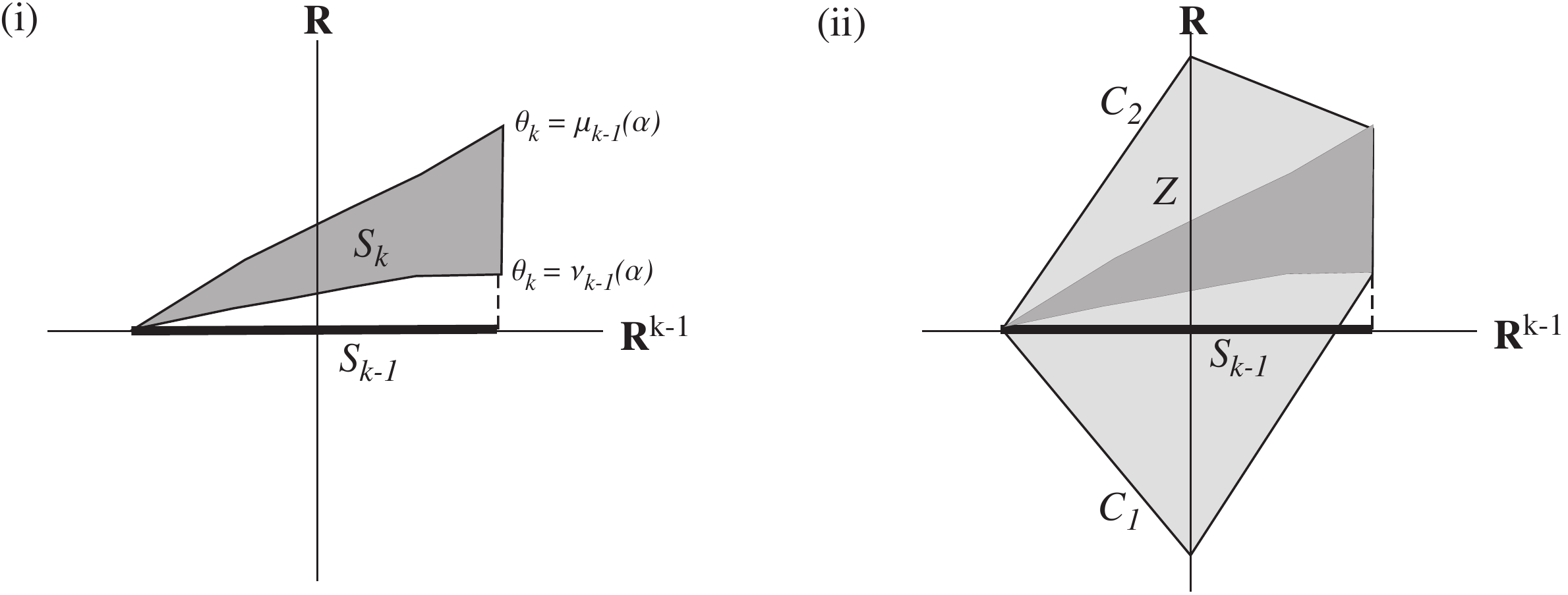}
\caption{(i) $\cs_k$ over $\cs_{k-1}$, (ii) region between two cones.}
\end{figure}

Let $Z$ be the region over $\cs_{k-1}$ lying between these two cones.  Then $\cs_k$ can be mapped homeomorphically onto $Z$ by taking each ``vertical" interval $\alpha \times R_k(\alpha)$ in $\cs_k$ and dilating it in the $\theta_k$-direction until it lies between the two cones.  Note that this dilation always makes sense when $\alpha \in \hbox{Int} \, \cs_{k-1}$ by Lemma 3, and no dilation is necessary when $\alpha \in \partial \cs_{k-1}$.  Finally, $Z$ is star-shaped in ${\bf R}^k$ with respect to the origin, so by radial projection it maps homeomorphically onto a closed $k$-dimensional ball, completing the proof.
\end{proof}

As mentioned earlier, a convex polygon is determined by its first $n-3$ turn angles, so $\xc \approx \cs$ is homeomorphic to $\cs_{n-3}$.  As a corollary to Proposition 4 we obtain the main result of this section:

\begin{thm} 
Assume that the configuration space $X(\vec{\ell})$ contains no straight line configurations.  Then the space $\xc$ of convex polygons with side lengths  $\vec{\ell}$ is homeomorphic to a closed Euclidean ball of dimension $n-3$.
\end{thm}

\section{Embedded configurations} 

We now determine the topology of the space $\xe$ of embedded configurations.  To do so, we use the following characterization of Euclidean space (see Brown [\textbf{2}] and Milnor [\textbf{11}]):

\medskip

\begin{assertion}
Let $M$ be an $n$-dimensional manifold such that every compact subset is contained in an open set homeomorphic to ${\bf R}^n$.  Then $M$ itself is homeomorphic to ${\bf R}^n$.
\end{assertion}

\medskip

To apply this to $\xe$, we proceed in two steps.  First we show that the subspace $\xc$ of convex configurations has an open neighborhood $U$ homeomorphic to ${\bf R}^{n-3}$.  Then given any compact subset $K$ of $\xe$, we adapt the techniques  of Cantarella, Demaine, Iben, and O'Brien [\textbf{4}] to stretch $U$ so that it covers $K$ by expanding it along the lines of gradient flow of a suitable ``energy" function.

\subsection{Putting a collar on the space of convex configurations}    
According to Theorem 5, the space $\xc$ of convex configurations is homeomorphic to a closed $(n-3)$-ball, which makes it reasonable to expect that $\xc$ can be thickened slightly within $\xe$ to obtain a neighborhood homeomorphic to an open ball, hence to ${\bf R}^{n-3}$.  We verify this by attaching a collar around $\xc$.  The existence of such a collar would follow from standard results in differential topology if $\xc$ were a smooth submanifold of $\xe$.  While the result here might also follow from general considerations, we give a direct argument instead that the singularities are mild enough that a collar may still be obtained.

A subspace $A$ of a topological space $X$ is said to be {\it bicollared} if there exists an embedding $h \colon A \times (-1, 1) \to X$ such that $h(a,0) = a$ for all $a$ in $A$.  Let $\partial \xc$ denote the boundary of $\xc$ in $\xe$.  It consists of convex configurations in which at least one turn angle is zero, and it separates the nonconvex configurations in $\xe$ from those convex configurations whose turn angles are all strictly positive.  By work of Brown [\textbf{3}, p.~337] and Connelly [\textbf{5}, p.~180], in order to prove that $\partial \xc$ is bicollared in $\xe$, it suffices to prove that it is bicollared locally.  Thus we need only find a suitable local model of how $\partial \xc$ sits inside $\xe$.  We do this by identifying local coordinates.  Recall that the dimension of $\xe$ is $n - 3$, so that's the number of coordinates we're looking for.

Let $a = (a_1, \dots , a_n)$ be a given polygon in $\partial \xc$ with turn angles $\alpha = (\alpha_1, \dots , \alpha_n)$.   These are to be regarded as fixed for the remainder of this section.  Any neighborhood of $a$ in $\xe$ contains nonconvex configurations, since some $\alpha_i$ is zero and there will be nearby configurations for which the corresponding turn angle $\theta_i$ is negative.  On the other hand, if $\alpha_j >0$, we may assume that all nearby configurations also satisfy $\theta_j > 0$.  The local picture depends on which of the $\alpha_i$ are zero.  We show that $n - 3$ of the turn angles, including those for which $\alpha_i = 0$, can be used to specify configurations in a neighborhood of $a$ uniquely.

Since $a$ is not a straight line configuration, at least three of its turn angles  are nonzero, say $\alpha_q, \alpha_r, \alpha_s \neq 0$ where $1 \leq q < r < s \leq n$.   In addition, $a$ is convex, so vertex $a_r$ does not lie on the line through $a_q$ and $a_s$.  Hence, the ordered triple of vertices $a_q, a_r, a_s$ has a well-defined orientation (e.g., clockwise or counterclockwise), and we may assume that, in all sufficiently close configurations $p = (p_1,  \dots, p_n)$, the triple $p_q, p_r, p_s$ has this same orientation as well. 

 We show that the $n-3$ turn angles $\theta_i$, $i \neq q, r, s$, work as local coordinates for $\xe$ near $a$, that is, a given set of  values $\theta_i \approx \alpha_i$ for these angles uniquely determines a configuration $p = (p_1, \dots , p_n)$ near $a$.  Certainly the given angles determine the three subchains from $p_q$ to $p_r$, from $p_r$ to $p_s$, and from $p_s$ to $p_q$, up to rotation and translation (Figure 8).  We must show that these subchains can be attached to one another  in only one way to form a polygon.   In fact, the subchain from $p_s$ to $p_q$ is completely determined since it contains the fixed segment $p_np_1$.   Also, the distances $|p_q - p_r|$, $|p_r- p_s|$, and $|p_s- p_q|$ are determined by the given angles, which gives two possible locations for $p_r$, but only one of these has the proper orientation of $p_q, p_r,  p_s$.  Hence, the entire configuration $p$ is uniquely determined, and we may parametrize a neighborhood of $a$ in $\xe$ by an open set $W$ in ${\bf R}^{n-3}$.

\begin{figure}[htb] 
\vskip -2 in
\hskip -.5 in
\includegraphics[width=7.5in]{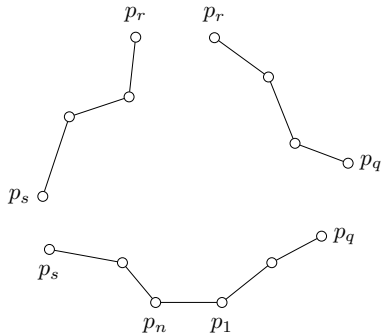}
\vskip -6.1 in
\caption{Three subchains determined by turn angles $\theta_i, i \neq q, r, s$.}
\end{figure}

As mentioned earlier, the local description of $\xc$ near $a$ depends on how many of the turn angles $\alpha_i$ are zero.  Up to homeomorphism, we may permute the coordinates in $W$ so that the coordinates for which $\alpha_i = 0$ come first.  Say there are $k$ such coordinates.  Then by a further change of variables, we may assume that $a$ corresponds to the point $(0, \dots , 0, \frac{1}{2}, \dots , \frac{1}{2})$  with an initial string of $k$ zeros and that $W$ itself is the open set
\[
W= (-1,1) \times \cdots \times (-1,1) \times (0,1) \times \cdots (0,1) = (-1,1)^k \times (0,1)^{n-3-k}.
\]
As far as the topology is concerned, the choice of intervals is arbitrary, but we are trying to draw a distinction between those coordinates for which the turn angle can be zero and those for which it cannot.  Recall that the ``missing" turn angles $\theta_q, \theta_r, \theta_s$ are also nonzero throughout a small enough neighborhood since we specifically selected $\alpha_q, \alpha_r, \alpha_s  \neq 0$.
Therefore, in these coordinates, the inclusion $\xc \subset \xe$ corresponds to 
\[
[0,1)^k \times (0,1)^{n-3 -k} \subset (-1,1)^k \times (0,1)^{n-3-k}.
\]
It is clear that the boundary of $[0,1)^k$ is bicollared in $(-1,1)^k$, hence the same remains true after taking the product with $(0,1)^{n-3-k}$.  This describes the local picture and shows that $\partial \xc$ is locally bicollared in $\xe$.  By the results of Brown and Connelly mentioned earlier, $\xc$ is bicollared in $\xe$.  Since $\xc$ is a closed $(n-3)$-ball, the union of $\xc$ with a collar is an open $(n-3)$-ball (Figure 9).    Thus we obtain the following result:

\begin{lem}
 $\xc$ has an open neighborhood $U$ in $\xe$ that is homeomorphic to ${\bf R}^{n-3}$.
 \end{lem}
 
 \begin{figure}[htb] \centering
\includegraphics[width=2.5in]{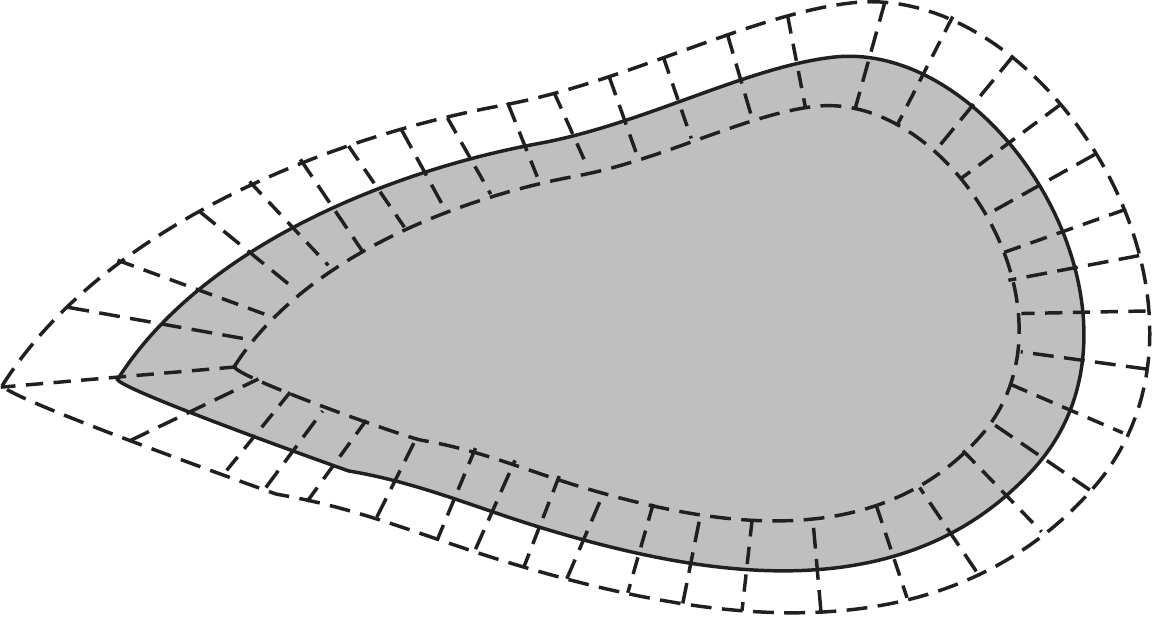}
\caption{Ball and collar.}
\end{figure}

For future reference, note that by shrinking $U$, if necessary, we may assume that the closure $\overline{U}$ in $\xe$ is compact.

\subsection{Reconfiguration along flow lines of vector fields}  

\renewcommand\theenumi{\roman{enumi}}
In [\textbf{4}], Canterella, Demaine, Iben, and O'Brien describe a method of convexifying a polygon by decreasing its ``energy."  For their purposes, energy is represented by a function $F \colon \xe \to {\bf R}$ having the properties that: 
\begin{enumerate}
\item $F$ is differentiable of class $C^2$ or higher; 
\item $F$ approaches $\infty$ as a polygon approaches self-intersection; and 
\item $F$ is decreasing to first order (i.e., strictly negative derivative) along strictly expansive motions.  (Recall that a motion in $\xe$ is called \textit{strictly expansive} if the distances between all pairs of vertices do not decrease and those distances between vertices not connected by a straight chain of edges actually strictly increase, again to first order.)  
\end{enumerate}

Connelly, Demaine, and Rote proved that given any nonconvex configuration $p$ in $\xe$ a strictly expansive motion through $p$ exists [\textbf{6}, pp.~214--227].  (On the other hand, at convex configurations, strictly expansive motions are forbidden by the Cauchy arm lemma.)  Hence, according to (iii) all critical points of $F$ lie in $\xc$.  This, together with (ii), means that following the direction of negative gradient flow in $\xe$ moves a polygon towards convexity.

Canterella et al.~exhibit a specific energy function satisfying (i)--(iii), which they call elliptic distance energy,   given by the formula:
\begin{equation}
F(p) = \sum_{\genfrac{}{}{0 pt}{}{\scriptstyle \hbox{\tiny edge  }\{p_i, p_{i+1}\}}{ \scriptstyle \hbox{\tiny vertex } p_j \neq p_i, p_{i+1} }} \frac{1}{ (| p_j - p_i| + |p_j - p_{i+1}  | - | p_{i+1} - p_i |)^2}.  
\end{equation}
The point is that the denominator of a typical summand vanishes if and only if $p_j$ lies on edge $p_ip_{i+1}$.  Thus $F$ is defined for all embedded polygons $p$.  Moreover, in order for a polygon to self-intersect, one of the vertices must approach one of the edges and hence $F$ goes to $\infty$.  For our purposes, however, it simplifies the argument to modify the energy slightly.  

Let $a \colon {\bf R} \to {\bf R}$ be a $C^\infty$ function such that 
\begin{equation*}
\begin{split}
a (x) = 0 &\hbox{\quad  if } x \leq 0, \\ \
 a(x) > 0 \>  \hbox{ and }  \> \frac{da}{dx} (x)  >0 &\hbox{\quad   if }x >0.  \\
 \end{split}
 \end{equation*}

A standard choice is to set $a (x) = e^{-1/x^2}$ when $x > 0$.  Now given an element $p = (p_1, \dots , p_n)$ in $\xe$ with turn angles $\theta = (\theta_1, \dots , \theta_n)$, define:
\begin{equation}
E(p) = \biggl(\sum_{\theta_i} a(-\theta_i)\biggr) \cdot F(p)
\end{equation}
where $F$ is the elliptic energy (2.2).  Note that only those turn angles for which $\theta_i < 0$ contribute to the sum.  Clearly $E$ still satisfies property (i).  To check (ii), suppose that $p$ approaches self-intersection.  Then at least one turn angle must approach a negative value.  (Otherwise, the configurations would remain convex, approaching a straight line configuration, which is not allowed.)  Therefore, the first factor in (2.3) is positive and bounded away from 0.  The second factor $F$ is known to approach $\infty$, and hence so does $E$.  Lastly suppose that $\alpha(t)$ is a strictly expansive motion.  This is necessarily through nonconvex configurations, so there is always at least one negative turn angle.  We denote derivatives with respect to time by $\dot{( \>\> )}$, write $\dot{E}$ to mean $\frac{d}{dt}(E \circ \alpha)$, and likewise for other such compositions.  Then
\begin{equation}
\dot{E} = \biggl(\sum_{\theta_i} - \frac{da}{d \theta}(-\theta_i)  \cdot \dot{\theta_i} \biggr) \cdot F + \biggl(\sum_{\theta_i} a(-\theta_i) \biggr) \cdot \dot{F}.
\end{equation}
In any strictly expansive motion, $\dot{\theta_i} < 0$ if $\theta_i >0$ and $\dot{\theta_i} > 0$ if $\theta_i  <0$. This follows from, say the law of cosines and the fact that in triangle $p_{i-1}p_ip_{i+1}$ the side $|p_{i+1} - p_{i-1}|$ is strictly increasing while the sides $\ell_i = |p_i - p_{i-1}|$ and $\ell_{i+1} = |p_{i+ 1} - p_i|$ are fixed.  Thus the first term in (2.4) is negative.  By the strictly expansive property applied to $F$, so is the second term.  Hence, $\dot{E} < 0$, showing that $E$ satisfies (iii).  This provides us with an energy function that is nonnegative on $\xe$ and zero precisely on the subset of convex configurations $\xc$.  In particular, by property (iii) $\xc$ is the set of critical points.

We use this to show that $\xe$ satisfies (2.1).  Let $K$ be a compact subset of $\xe$.  By replacing $K$ with $K \cup \overline{U}$,  where $U$ is the open set of Lemma 6, we may assume that $\overline{U }\subset K$.  (Recall that  $\overline{U}$ can be taken to be compact.)  The basic idea is to let $U$ grow outwards under the effect of the gradient of $ E$.  Perhaps it is somewhat simpler first to reverse things and to think of $K$ flowing towards $\xc$ under the negative gradient.  This backwards flow reduces the energy of the points of $K$ and brings them towards convexity, so after some finite amount of time $\tau$, $K$ is shrunk inside $U$.  We would then like to say that, going forward in time, it follows that under positive gradient flow $U$ is stretched so that it covers $K$  after time $\tau$.    The stretched set would satisfy (2.1).  It is certainly true that some portion of $U$ covers $K$ after time $\tau$.  But it is possible that some points of $U$ fly off to infinity before time $\tau$, so that the gradient flow is not defined on all of $U$ over the entire interval $[0,\tau]$.  As a result, we make some technical adjustments to the vector field along which the flow takes place.

\begin{figure}[htb] \centering
\includegraphics[width=4in]{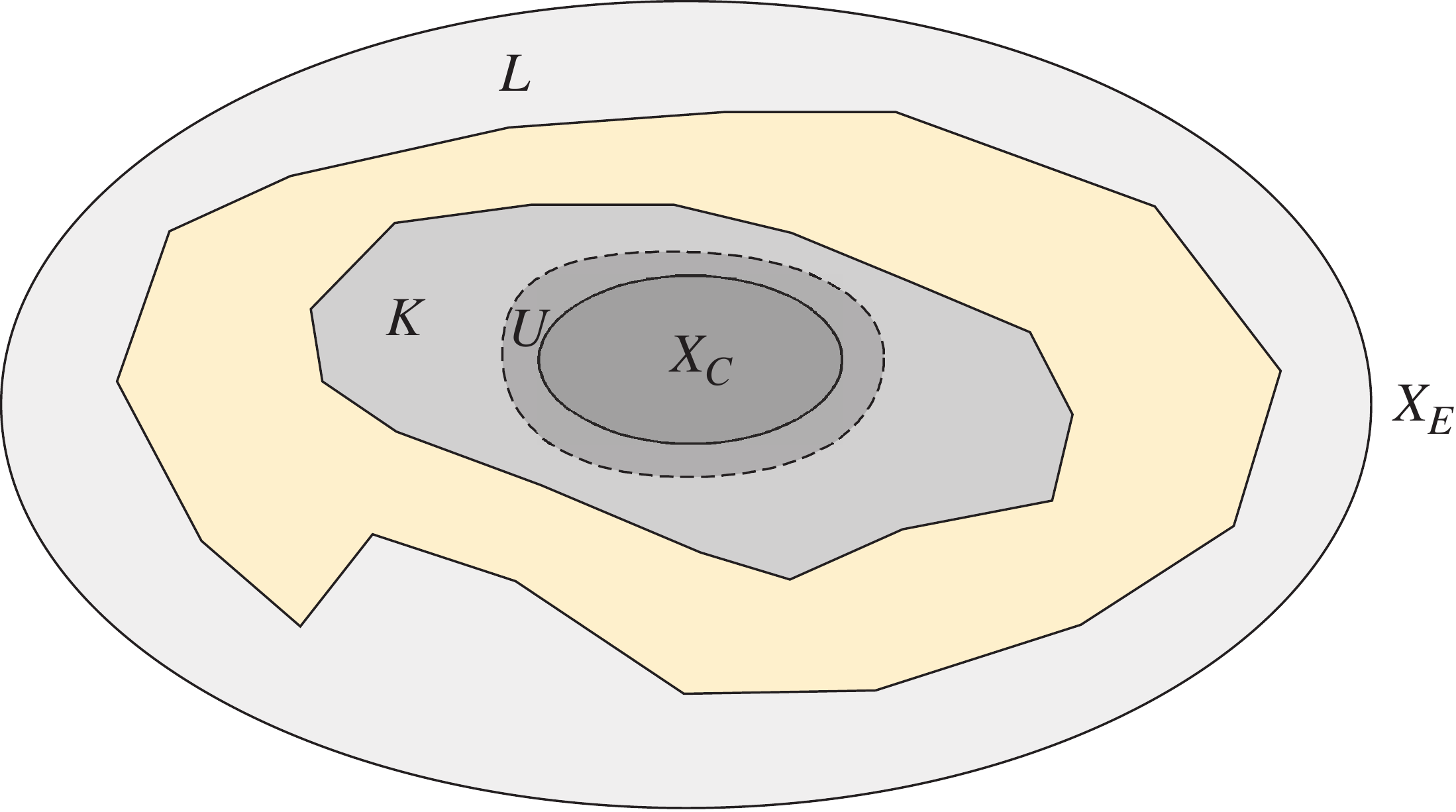}
\caption{$\xc \subset U \subset K, K \cap L = \emptyset$.  $U$ is stretched until it covers $K$.}
\end{figure}

Let $m = \max_{p \in K} E(p)$, and  let $L = \{ p \in \xe \; | \; E(p) \geq m + 1 \}$.    Then $L$ is a closed subset of $\xe$ such that $K \cap L = \emptyset$ (Figure 10). Using a partition of unity (which in this case reduces to the $C^\infty$ Urysohn lemma), there exists a $C^\infty$ function $b \colon \xe \to [0,1]$ such that $b =1$ on $K$ and $b = 0$ on $L$.  Consider the vector field $\xi = b\cdot \hbox{grad}\, E$.  It agrees with $\hbox{grad}\,E$ on $K$ and vanishes on $L$.  Let $\varphi_t$ denote the flow associated to $\xi$.  By the arguments given in the preceding paragraph, there exists a finite time $\tau > 0$ such that $\varphi_\tau(U)$ covers $K$.  Thus $\varphi_\tau(U)$ is an open neighborhood of $K$ homeomorphic to $U$, which is homeomorphic in turn to ${\bf R}^{n-3}$.  By (2.1) this proves the following result:

\begin{thm} Assume that the configuration space $X(\vec{\ell})$ contains no straight line configurations.  Then the space $\xe$ of embedded polygons with side lengths $\vec{\ell}$ is homeomorphic to the Euclidean space ${\bf R}^{n-3}$.
\end{thm}

\section{A noncontractible closure}

As a brief final note, we give an example in which the space $\xe$ of embedded polygons is contractible, but its closure $\overline{X}_E$ is not.  This gives a negative resolution to a conjecture posed by Connelly, Demaine, and Rote [\textbf{6}, p.~235].

For the example, consider quadrilaterals with side lengths $\vec{\ell} = (6, 4, 2, 4)$.  The main point is that all configurations of these quadrilaterals are embeddings with one exception, which occurs when the edges are folded over to lie along a line.

The full configuration space $X$ is homeomorphic to a figure eight.  One of the lobes consists of all the counterclockwise embeddings together with a straight line configuration which represents the point at which the two lobes are attached.  This is illustrated in Figure 11.  (The second lobe consists of the reflections of the first in the side of length 6.)  Thus $\xe$ is homeomorphic to a circle with a point deleted, which is contractible, while $\overline{X}_E$ is homeomorphic to a circle, which is not.

\begin{figure}[htb] \centering
\vskip -1 in
\includegraphics[width=4 in]{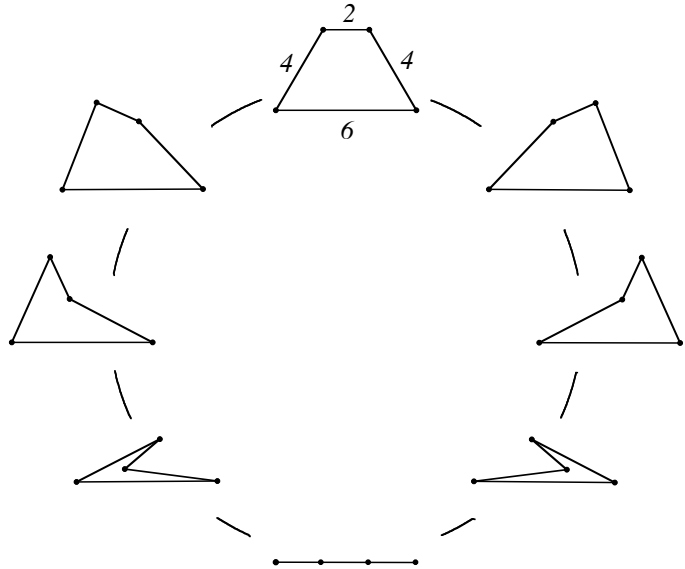}
\vskip -1 in
\caption{A circle of quadrilaterals with side lengths $6,4,2,4$.}
\end{figure}

Of course, this example involves a straight line configuration, which we have been excluding until now.  Whether there is an example in which straight line configurations are not allowed remains to be seen.

\end{document}